\documentclass[a4paper,11pt]{article}

\pdfoutput=1 

\usepackage{./style_files/jheppub} 

\usepackage[T1]{fontenc} 

\title{\boldmath Fisher information as a probe of spacetime structure: Relativistic quantum metrology in (A)dS}

\author[a,b,c,1]{Haoxing Du,\note{Corresponding author.}}
\author[a,b]{Robert B. Mann}

\affiliation[a]{Perimeter Institute for Theoretical Physics, \\ 31 Caroline St N, Waterloo, Ontario, Canada N2L 2Y5}
\affiliation[b]{Department of Physics and Astronomy, University of Waterloo, \\ Waterloo, Ontario, Canada N2L 3G1}
\affiliation[c]{Department of Physics, University of California, Berkeley, \\ Berkeley, California, USA 94720}

\emailAdd{haoxing\_du@berkeley.edu}
\emailAdd{rbmann@uwaterloo.ca}

\abstract{Relativistic quantum metrology studies the maximal achievable precision for estimating a physical quantity when both quantum and relativistic effects are taken into account.
We study the relativistic quantum metrology of temperature in $(3+1)$-dimensional de Sitter and anti-de Sitter space. 
Using Unruh-DeWitt detectors coupled to a massless scalar field as probes and treating them as open quantum systems, we compute the Fisher information for estimating temperature. 
We investigate the effect of acceleration in dS, and the effect of boundary condition in AdS.
We find that the phenomenology of the Fisher information in the two spacetimes can be unified, and analyze its dependence on temperature, detector energy gap, curvature, interaction time, and detector initial state. 
We then identify estimation strategies that maximize the Fisher information and therefore the precision of estimation.}

\usepackage{braket}
\usepackage{color}
\usepackage[dvipsnames]{xcolor}

\newcommand{\D}[2]{\frac{\partial #1}{\partial #2}}

\renewcommand{\vec}[1]{\boldsymbol{#1}}
\newcommand{\F}{\mathcal{F}}
\newcommand{\G}{\mathcal{G}}

\usepackage[utf8]{inputenc}
\usepackage[T1]{fontenc}
\usepackage{fancyhdr}
\usepackage{lipsum}
\usepackage{graphicx}
\usepackage{titlesec}
\usepackage{appendix}
\usepackage[sectionbib]{chapterbib}
\usepackage[breakwords]{truncate}
\usepackage{lastpage}
\usepackage{amsmath}
\usepackage{hyperref}
\usepackage{amssymb}
\usepackage[font={small}]{caption}
\usepackage{physics}

\begin{document}
\maketitle
\flushbottom

\newpage
\section{Introduction} \label{ch:intro}

The reconcilation of quantum theory and general relativity  remains one of the greatest unsolved problems in physics today. In recent years, numerous insights have come from studying quantum information theory in the context of quantum field theory in flat and curved spacetime, a subdiscipline now known as \emph{Relativistic Quantum Information} (RQI) \cite{mann2012relativistic}. By looking closely at the behavior of quantum systems in the presence of relativistic effects, we gain insight on the interplay between quantum theory and relativity, which is an essential step towards a complete theory of quantum gravity.

At a much more pragmatic level,   quantum technologies are rapidly becoming a reality, and soon   will be applied in settings in which relativistic effects cannot be neglected. It is therefore just as crucial to understand both relativistic and gravitational effects on various quantum technologies, including quantum communication, quantum computing, and quantum metrology. This latter subdiscipline refers to the study of techniques that exploit quantum resources to perform measurements to a higher precision than is possible classically \cite{giovannetti2011advances}. For example, it has been shown that quantum squeezed states can be used to enhance the precision of interferometers \cite{Caves1981quantum}, including those used to detect gravitational waves \cite{grote2013first}. Likewise, quantum entanglement has been shown to be capable to improve measurement precision in the presence of noise \cite{demkowicz2014using}.


In recent years, there has been considerable interest in applying quantum metrological techniques in relativistic settings, specifically, to relativistic quantum fields in curved spacetimes \cite{ahmadi2014quantum}. The precision of certain measurements can be further enhanced by exploiting relativistic effects on quantum systems \cite{ahmadi2014relativistic}, a line of research  termed \emph{relativistic quantum metrology}. It is especially interesting when the quantity being measured is a parameter governing some phenomenon in which both quantum and gravitational effects are present. Previous research has investigated the metrology of a wide range of relativistic phenomena, including acceleration \cite{ahmadi2014relativistic}, the Unruh-Hawking effect \cite{aspachs2010optimal, borim2020precision}, the expansion rate of a Robertson-Walker universe \cite{wang2015parameter}, Bose-Einstein condensates \cite{Sabin2014,Schutzhold2018,Robbins2019},
and Schwarzschild spacetime parameters \cite{bruschi2014quantum}.

In many applications, since the quantity being measured does not correspond to a quantum observable, the measurement becomes a parameter estimation problem, whose key
figure of merit is the  Fisher information.  This is the quantity that measures how sensitive the state of a system is to a change in the parameter of interest. As a consequence of the Cram\'er-Rao bound \cite{cramer1999mathematical, rao1992information}, Fisher information is also the quantity that determines the maximal precision that can be possibly achieved in a metrological task.

In this paper we carry out an investigation of relativistic quantum metrology in curved spacetime, restricting our attention to the constant-curvature de Sitter (dS) and Anti-de Sitter (AdS) cases. Making use of Unruh-DeWitt detectors as probes of the quantum vacuum,
our work extends an earlier study \cite{huang2018quantum} of Fisher information as a probe for estimating the expansion rate (i.e. the Hubble parameter) with a comoving detector in different dS vacua.  We find that the behavior of the Fisher information is somewhat more complicated than that presented in Ref. \cite{huang2018quantum}. We show that   this framework can be used to estimate the KMS temperature of the thermal response the detector, and examine the effect of acceleration in dS and the effect of boundary condition in AdS. We find that the behavior of the Fisher information in dS and AdS share similar phenomenology, particularly at late times. We also study the dependence of the Fisher information on various parameters in both spacetimes, and identify strategies for maximizing it.

\section{General Formalism}

\subsection{Fisher Information}

The  \emph{parameter estimation problem}
is concerned with how the value of a given physical quantity that is not directly observable can be inferred via its statistical effects on quantities that are observable. Consider  an observable quantity $x$, which  is sampled from a family of probability distributions $p(x|\xi)$, labeled by a parameter  $\xi$ that is the quantity of interest.
The parameter estimation problem is that of estimating $\xi$ from the data set of $x$.

Since the data set is inevitably has noisy, any estimate of $\xi$ will contain some error. The   \emph{Cram\'er-Rao bound} provides a bound on how well the estimation can be performed. Letting $(\chi,\Xi)$ be the respective sets of values that $x$ and $\xi$ can take, an \emph{estimator} for $\xi$ for a sample size $n$ is a function $\hat{\xi}: \chi^n \to \Xi$. The estimator is \emph{unbiased} if its expected value is equal to the true value of the parameter. The \emph{Fisher information} 
\begin{equation}\label{eq:fisher-def}
    \F(\xi) = \int p(x|\xi) \left( \D{\ln p(x|\xi)}{\xi} \right)^2 dx = \int \frac{1}{p(x|\xi)} \left( \D{p(x|\xi)}{\xi} \right)^2 dx
\end{equation}
quantifies the amount of information that the observable $x$ carries about   $\xi$. The Cram\'er-Rao bound 
\begin{equation}\label{eq:crbound}
    \text{var}(\hat\xi) \geq \frac{1}{\F(\xi)}
\end{equation}
states that the mean-squared error of any unbiased estimator $\hat\xi$ of   $\xi$ is lower bounded by the reciprocal of the Fisher information.
In other words, Fisher information is the quantity that yields an ultimate bound on the precision attainable in a parameter estimation problem.

\subsection{de Sitter and Anti-de Sitter space}

Both dS$_4$ and AdS$_4$ can be represented as 4-dimensional hyperboloids embedded in 5-dimensional spacetimes,
\begin{equation}\label{eq:minkowski5}
    ds^2= - dX_0^2 + dX_1^2 + dX_2^2+ dX_3^2 \pm dX_4^2,
\end{equation} 
with dS$_4$ taking the plus sign. (A)dS$_4$ can be obtained as the hyperboloid
\begin{equation}\label{eq:ds-hyp}
     - X_0^2 + X_1^2 + X_2^2 + X_3^2 \pm X_4^2 = \pm \ell^2,
\end{equation}
with dS$_4$ again taking the plus signs. The well-known static (A)dS metric
\begin{equation}\label{eq:ds-static}
    ds^2 = - \left( 1 \pm \frac{r^2}{\ell^2} \right) dt^2 + \left( 1 \pm \frac{r^2}{\ell^2} \right)^{-1}dr^2 + r^2d\Omega_2^2,
\end{equation}
is obtained via the transformations
\begin{equation}
    X_1 = r\sin\theta\sin\phi, \quad
    X_2 = r\sin\theta\cos\phi, \quad
    X_3 = r\cos\theta
\end{equation}
and
\begin{align}\label{eq:ds-static-trsf}
    X_0 = \sqrt{\ell^2-r^2} \sinh(t/\ell), \quad
    X_4 = \sqrt{\ell^2-r^2} \cosh(t/\ell)
\end{align}
\begin{align}\label{eq:ads-static-trsf}
    X_0 = \sqrt{\ell^2+r^2} \sin(t/\ell), \quad
    X_4 = \sqrt{\ell^2+r^2} \cos(t/\ell)
\end{align}
for dS$_4$ and AdS$_4$ respectively. The quantity $\ell =\sqrt{3/\Lambda}\equiv 1/k$ is referred to as the (A)dS length in both cases, the dS case (which takes the minus sign in Eq.~\eqref{eq:ds-static}) having a coordinate singularity at the \emph{cosmological horizon} $r = \ell$.   

For dS$_4$ we shall also make use of \emph{comoving coordinates}, with metric
\begin{equation}\label{eq:ds-flat}
    ds^2 = -dt^2 + e^{2t/\ell} (dx_1^2 + dx_2^2 + dx_3^2),
\end{equation}
obtained from the flat space embedding via the transformations
\begin{align}\label{eq:ds-flat-trsf}
    T &= \ell \sinh(t/\ell) + \frac{r^2 e^{t/\ell}}{2\ell},
    &X_0 &= \ell \cosh(t/\ell) - \frac{r^2 e^{t/\ell}}{2\ell}, \nonumber \\
    X_1 &= e^{t/\ell}x_1,
    &X_2 &= e^{t/\ell}x_2,
    &X_3 &= e^{t/\ell}x_3.
\end{align}

We study a massless real scalar field $\phi$ conformally coupled to curvature via the action
\begin{equation}\label{eq:scalact}
    S = \int \sqrt{-g} \left[\frac{1}{2} g^{\mu\nu}\nabla_\mu \phi \nabla_\nu \phi - \frac{1}{12} R \phi^2\right].
\end{equation}
We work in the conformal vacuum, since it is the natural vacuum that obeys the symmetry of the de Sitter group \cite{birrell1984quantum}. The Wightman function $W(x,x') = \ev{\phi(x)\phi(x')}{0}$ for a comoving observer in the conformal vacuum of dS$_4$ is known to be
\begin{equation}\label{eq:wightman-ds}
    W^{\text{dS}}(x(\tau),x(\tau')) =-\frac{1}{2\pi\ell\sqrt{2}}  \left(\frac{1}{\sinh^2\left(\frac{\tau - \tau'}{\ell}  -i\epsilon\right)}  \right).
\end{equation}
Note that this is a function of $\Delta\tau = \tau - \tau'$ only, implying that the comoving trajectory is stationary.

Defining a quantum field theory in anti-de Sitter space involves a lot more subtleties, because AdS space is not globally hyperbolic, i.e. it admits no Cauchy surface. The standard approach is to circumvent this problem by first quantizing the field in the Einstein Static Universe and then translating the result back to AdS space, yielding~\cite{avis1978quantum}
\begin{equation}\label{eq:wightman}
    W^\text{AdS}(x,x')=\frac{1}{4\pi\ell\sqrt{2}} \left(\frac{1}{ {\sigma(x,x')}}-\frac{\zeta}{ {\sigma(x,x')+2}}\right),
\end{equation}
for the Wightman function,
where $2\ell^2\sigma(x,x')$  is the geodesic distance between $x$ and $x'$, and $\zeta = -1,0,$ or 1 specifies either Dirichlet ($\zeta=1$), transparent ($\zeta=0$), or Neumann ($\zeta=-1$) boundary conditions. The geodesic distance $\sigma(x,x')$ is straightforwardly computed from the embedding space Eq.~\eqref{eq:minkowski5}.

\subsection{Unruh-DeWitt detectors} \label{sec:udw}

We employ the Unruh-DeWitt model of particle detectors \cite{unruh1976notes, dewitt1979quantum}, in which a two-level quantum system, whose two energy levels $\ket{0}_D$ and $\ket{1}_D$ are separated by an energy gap $\Omega$. Known as an \emph{Unruh-DeWitt (UDW) detector}, 
it moves along a trajectory $x(\tau)$ parametrized by proper time $\tau$, and interacts with the massless scalar field $\phi(x)$, via the interaction Hamiltonian \footnote{Compared to the most general UDW interaction, we consider only point-like detectors and no switching function.}
\begin{equation}\label{eq:udw-ham}
    H_I = \lambda \left( e^{i\Omega\tau} \sigma^+ + e^{-i\Omega\tau} \sigma^- \right) \otimes \phi[x(\tau)].
\end{equation}
where $\lambda$ is a constant that controls the strength of the interaction, and $\sigma^+ = \ket{1}_D\bra{0}_D$ and $\sigma^- = \ket{0}_D\bra{1}_D$ are ladder operators on the Hilbert space associated with the detector.

To lowest order in $\lambda$, the probability of transition 
from the ground state $\ket{0}_D$ to the excited state $\ket{1}_D$ is 
proportional to a quantity called the \emph{response function} \cite{louko2008transition, jennings2010response},
\begin{equation}
    G(\omega) = \int_{-\infty}^\infty d\tau \int_{-\infty}^\infty d\tau'\, e^{-i\omega(\tau-\tau')}\, W(x(\tau),x(\tau')),
\end{equation}
where $W(x(\tau),x(\tau'))$ is the Wightman function.  Trajectories for which $W(x(\tau),x(\tau'))$ becomes  a function only of $\Delta\tau = \tau - \tau'$, are called  \emph{stationary} trajectories. In this case, it is often more useful to consider the \emph{response per unit time}  \cite{louko2008transition, jennings2010response},
\begin{equation}\label{eq:detec-response}
    \mathcal{G}(\omega) = \int_{-\infty}^\infty d\Delta \tau \; e^{-i\omega\Delta\tau}\, W(\Delta\tau)
\end{equation}
a quantity that plays a central role when considering UDW detectors as open quantum systems, as we shall see.

The key utility of UDW detectors is that they provide an operational   interpretation for thermality, and more generally, for discerning the structure of the quantum vacuum.
Thermal states arise when event horizons are present \cite{sewell1982quantum}, such as in the expanding de Sitter spacetime \cite{ gibbons1977cosmological, birrell1984quantum}, in the presence of a black hole \cite{hawking1975particle}, and for uniform detector accelerations. In AdS space, a UDW detector will experience thermal radiation provided the acceleration is sufficiently large \cite{deser1997accelerated,jennings2010response}. 
 
We are interested in the thermal response of a UDW detector in three scenarios: (a) a comoving detector in dS$_4$, (b) a uniformly accelerated detector in dS$_4$, and (c) a uniformly accelerated detector in AdS$_4$. Fortunately, for these three cases of our interest, the integral Eq. \eqref{eq:detec-response} can be evaluated analytically and the results are known.

\subsubsection{Comoving detector in dS$_4$}


The response of a comoving detector in de Sitter space is well-known \cite{birrell1984quantum}
\begin{equation}
    \mathcal{G}^{\text{dS}}(\omega) = \frac{\omega}{2\pi} \frac{1}{e^{2\pi\ell\omega}-1},
\end{equation}
and can be obtained from directly evaluating the Fourier transform of the Wightman function Eq.~\eqref{eq:wightman-ds} with contour integration. Note the presence of the Planck factor $1/(e^{2\pi\ell\omega}-1)$, which indicates that this is a thermal spectrum with temperature
\begin{equation}
    T = \frac{1}{2\pi\ell}.
\end{equation}
This fact that a comoving observer in de Sitter space experiences a thermal bath of radiation is known as the \emph{Gibbons-Hawking effect}, and the above temperature is sometimes referred to as the Gibbons-Hawking temperature \cite{gibbons1977cosmological}. Note that the response only depends on the spacetime curvature $\ell$ through this temperature, and the temperature is determined by curvature alone.   We can write the response per unit time in terms of $T$
\begin{equation}\label{eq:ds-response}
    \mathcal{G}^{\text{dS}}(\omega) = \frac{\omega}{2\pi} \frac{1}{e^{\omega/T}-1}
\end{equation}
instead of the de Sitter length $\ell$.

\subsubsection{Accelerated detector in dS$_4$}

A uniformly accelerated detector in dS$_4$ moves along the trajectory
\begin{equation}
    r e^{t/\ell} = K
\end{equation}
using  the coordinates Eq.~\eqref{eq:ds-flat},
where $K$ is a constant, and
has a response per unit time  
\cite{casadio2011unruh}
\begin{equation}
    \mathcal{G}^\text{dS}_\text{accel} = \frac{\omega}{2\pi} \frac{\sqrt{1-K^2/\ell^2}}{\exp\left(2\pi\ell\sqrt{1-K^2/\ell^2} \,\omega\right)-1}.
\end{equation}
where the temperature \cite{narnhofer1996hot,deser1997accelerated}
\begin{equation}
    T = \frac{1}{2\pi\ell\sqrt{1-K^2/\ell^2}} = \frac{\sqrt{a^2 + 1/\ell^2}}{2\pi}.
\end{equation}
can   be recognized from the Planck factor.  The magnitude of the 4-acceleration is
\begin{equation}
    a^2 = - \frac{K^2}{\ell^2(\ell^2-K^2)}.
\end{equation}
Note that the form of this temperature is suggestive of contributions from both the Unruh effect (the $a^2$ term) and the Gibbons-Hawking effect (the $1/\ell^2$ term).

We can again write the response per unit time in terms of the temperature, obtaining
\begin{equation}\label{eq:ds-response-a}
    \mathcal{G}^\text{dS}_\text{accel} = \frac{\omega}{4\pi^2} \, \frac{1}{\ell T} \, \frac{1}{e^{\omega/T}-1}.
\end{equation}
In the limit $a \to 0$, $T \to 1/2\pi \ell$, and we recover the comoving response Eq.~\eqref{eq:ds-response}.

\subsubsection{Accelerated detector in AdS$_4$}

A uniformly accelerated detector in AdS space also experiences a nonzero temperature \cite{deser1997accelerated}, and has a response \cite{jennings2010response} \begin{equation}\label{eq:ads-response-a}
    \mathcal{G}^{\text{AdS}}(\omega) = \left( \frac{\omega}{2\pi} -  \frac{\zeta}{4 \pi a\ell^2} \sin\left[ \frac{2\omega\ell}{\sqrt{a^2\ell^2-1}} \sinh^{-1} \left( \sqrt{a^2\ell^2-1} \right) \right] \right) \frac{\Theta(a\ell-1)}{\exp\left( \frac{2\pi \omega\ell}{\sqrt{a^2\ell^2-1}} \right) -1},
\end{equation}
where $\Theta(\cdot)$ is the Heaviside function, and $\zeta=-1,0,1$ specifies the boundary conditions\footnote{
The expression \eqref{eq:ads-response-a} is equivalent to that in  \cite{deser1997accelerated} 
upon identifying
$b = \cosh^{-1}(1 + 2a^2 \ell^2)$. (The AdS length $\ell$ is denoted as $R$ in Ref.~\cite{deser1997accelerated}.)}. Note that the first term again corresponds to a thermal spectrum, with temperature
\begin{equation}
    T = \frac{\sqrt{a^2\ell^2-1}}{2\pi\ell}.
\end{equation}
 In contrast with an accelerated detector in flat space, which experiences a temperature $T = a/2\pi$ for all nonzero acceleration $a$, the detector in AdS$_4$ has a nonzero response only when $a$ is greater than a certain critical acceleration $k = 1/\ell$.  
 In other words, there is no thermal response unless the acceleration $a$ exceeds the inverse of the AdS length.
  
The second term in \eqref{eq:ads-response-a}
gives a  correction to the thermal spectrum of magnitude $1/a\ell^2$, depending on the boundary condition. Again, we can write the response in terms of temperature $T$, obtaining
\begin{equation}\label{eq:ads-response}
    \mathcal{G}^{\text{AdS}}(\omega) = \left( \frac{\omega}{2\pi} -   \frac{\zeta}{4 \pi\ell \sqrt{4\pi^2T^2\ell^2 + 1}} \sin\left[ \frac{\omega}{\pi T} \sinh^{-1} \left( {2\pi T\ell} \right) \right] \right) \frac{\Theta(T)}{e^{\omega/T}-1}.
\end{equation}
Note that when $\zeta = 0$, i.e. when transparent boundary condition is taken, Eq.~\eqref{eq:ads-response} becomes identical to the response in dS$_4$ for a \emph{comoving} observer, Eq.~\eqref{eq:ds-response}. One might naively have expected that an  accelerated detector in either spacetime will  experience a similar effect. We can make sense of this by recalling that there is only one ``source of thermality'' for both a comoving detector in dS  (the source being the cosmological horizon) and an accelerated detector in AdS  (the source being acceleration), whereas both of these sources contribute to the thermality of an accelerated observer in dS.

As we will see below, the motion of the detector only comes into play for the Fisher information through the response per unit time. Consequently the Fisher information for   an accelerated detector in AdS  with transparent boundary conditions is exactly the same as that of a comoving detector in dS. In our discussion below on the effect of different AdS boundary conditions on the Fisher information, we implicitly assume that the AdS case with transparent boundary conditions also represents a comoving detector in dS.

\section{Metrology with Unruh-DeWitt detectors} \label{ch:udw}

We shall consider UDW detectors as open quantum systems \cite{huang2018quantum, benatti2004entanglement},  motivated by the fact that a detector experiencing thermal radiation behaves exactly like a system immersed in an external heat bath. Any such interaction can be studied by considering the time evolution of the overall system, and then tracing out the degrees of freedom belonging to the environment, to obtain the dynamics of the system of interest. The resultant dynamics of the system is in general non-unitary, and information is lost to the environment via their interaction.  

Here we treat the UDW detector as the ``system'', and the fluctuating, noisy vacuum of the quantum field as the ``environment''. Full dynamics of the detector is obtained by tracing over the field degrees of freedom, and as we will see shortly, has a solvable form when specific assumptions are made. This framework has been applied previously to study the Fisher information for metrological tasks in flat space \cite{tian2015relativistic, hao2016quantum}. 

We start by considering the overall Hamiltonian of the combined system that includes the detector and the quantum field,
\begin{equation}
    H = H_D + H_\phi + H_I,
\end{equation}
where $H_D = \frac12 \Omega a_D^\dagger a_D = \frac{1}{2}\Omega(\ket0_D\bra0_D - \ket1_D\bra1_D)$ is the free Hamiltonian of the detector with energy gap $\Omega$, $H_\phi= \sum_\mathbf{k} \omega_\mathbf{k} a_\mathbf{k}^\dagger a_\mathbf{k}$ is the free Hamiltonian of the massless scalar field $\phi(x)$, and $H_I$ is the Unruh-DeWitt interaction Hamiltonian given by Eq. \eqref{eq:udw-ham}.

The time evolution of the combined system is governed by the von Neumann equation
\begin{equation}
    \D{\rho_\text{tot}}{\tau} = -i [H, \rho_\text{tot}],
\end{equation}
where $\rho_\text{tot}$ is the density matrix of the combined system, which starts in the initial state $\rho_\text{tot}(0) = \rho_D(0) \otimes \ket{0}\bra{0}$, where $\rho_D(0)$ is the initial state of the detector, and $\ket{0}$ is the conformal vacuum of the scalar field $\phi(x)$.

The state of the detector can be obtained by taking the partial trace over the field of the combined state, i.e. $\rho_D = \tr_\phi \rho_\text{tot}$. It turns out that when the coupling is weak ($\lambda \ll 1$), and the field correlations decay fast enough at large time separations, the evolution of the density operator of the detector is captured by the \emph{master equation of Kossakowski-Lindblad form} \cite{benatti2004entanglement}. This is the most general description of Markovian time evolution of a quantum system \cite{manzano2020short}, and is given by
\begin{equation}\label{eq:lindblad}
    \D{\rho_D(\tau)}{\tau} = -i[H_\text{eff}, \rho_D(\tau)] + L[\rho_D(\tau)],
\end{equation}
where $H_\text{eff} = \frac12\Tilde{\Omega}(\ket0_D\bra0_D - \ket1_D\bra1_D)$, and
\begin{equation}
    L[\rho] = \frac12 \sum_{i,j=1}^3 C_{ij} \left( 2\sigma_j\rho\sigma_i - \sigma_i\sigma_j\rho - \rho\sigma_i\sigma_j\right).
\end{equation}
Here, $\sigma_i$ are the Pauli matrices, and $\Tilde{\Omega}$ is a renormalized gap given by 
\begin{equation}
    \Tilde{\Omega} = \Omega + i \left[ \mathcal{K}(-\Omega) - \mathcal{K}(\Omega) \right],
\end{equation}
where $\mathcal{K}(\Omega)$ is the Hilbert transform of the response per unit time $\mathcal{G}(\omega)$ defined by
\begin{equation}
    \mathcal{K}(\Omega) = \frac{1}{i\pi} \text{ PV} \int_{-\infty}^\infty d\omega \, \frac{\mathcal{G}(\omega)}{\omega - \Omega},
\end{equation}
where PV denotes the Cauchy principal value. $C_{ij}$ is called the \emph{Kossakowski matrix}, and is also completely determined by the response per unit time $\mathcal{G}(\omega)$:
\begin{equation}
    C_{ij} = \begin{pmatrix}
    A & -iB & 0 \\
    iB & A & 0 \\
    0 & 0 & A+C
    \end{pmatrix},
\end{equation}
where
\begin{equation}\label{eq:A}
    A = \frac12[\mathcal{G}(\Omega) + \mathcal{G}(-\Omega)],
\end{equation}
\begin{equation}\label{eq:B}
    B = \frac12[\mathcal{G}(\Omega) - \mathcal{G}(-\Omega)],
\end{equation}
\begin{equation}
    C = \mathcal{G}(0) - A.
\end{equation}

Somewhat miraculously, Eq. \eqref{eq:lindblad} can be solved analytically. For a detector initialized in the general pure state $\cos\frac{\theta}{2} \ket{0} + \sin\frac{\theta}{2} \ket{1}$, its density matrix at time $\tau$ is specified by the Bloch vector $\vec{a} = (a_1,a_2,a_3)$ such that
\begin{equation}
    \rho(\tau) = \frac12 \left( I + \vec{a}(\tau) \cdot \vec{\sigma}  \right),
\end{equation}
where $\vec{\sigma} = (\sigma_1,\sigma_2,\sigma_3)$ are the Pauli matrices, and the Bloch vector components are given by
\begin{equation}
    a_1(\tau) = e^{-A\tau/2} \sin\theta\cos\Tilde{\Omega}\tau,
\end{equation}
\begin{equation}
    a_2(\tau) = e^{-A\tau/2}\sin\theta\sin\Tilde{\Omega}\tau,
\end{equation}
\begin{equation}\label{eq:az}
    a_3(\tau) = -e^{-A\tau} \cos\theta - R(1-e^{-A\tau}).
\end{equation}
Here, $\Tilde{\Omega}$ is the renormalized frequency defined above, and $R = B/A$. Note that $|\vec{a}| < 1$ in general, implying that the evolution is non-unitary.

This approach is more general than simply looking at the probability of transition and the response function, since it allows us to access the time evolution of the state of the UDW detector. As we will see in Section \ref{ch:results}, access to the detailed time evolution of the detector gives us enhancement of the Fisher information. By taking this approach, we also avoid troubles of regularization, and do not fall prey to effects that are only present for certain types of switching functions \cite{louko2008transition}.

Next, we compute the Fisher information for estimating temperature in dS  and AdS with UDW detectors. The estimation strategy we take is simply that of letting the detector interact with the massless scalar field in the spacetime of interest, and then making a projective measurement of the detector's state after some time $\tau$. In this case, since the Hilbert space associated with the UDW detector has dimension two, every measurement has two possible outcomes. Recalling the definition of Fisher information, Eq. \eqref{eq:fisher-def}, we see that the probability distribution $p(x|\lambda)$ reduces to the probability of getting either outcome, and the integral simplifies to a sum of two terms. 
Suppose the two outcomes occur with probabilities $p$ and $1-p$. Then the Fisher information takes the simpler expression
\begin{equation}
    \F(\lambda) = \frac{1}{p} \left( \D{p}{\lambda} \right)^2 + \frac{1}{1-p} \left( -\D{p}{\lambda} \right)^2 = \frac{1}{p(1-p)} \left( \D{p}{\lambda} \right)^2.
\end{equation}

Consider measuring the detector in the computational basis 
$\{ \ket{0}_D, \ket{1}_D \}$. For a detector in the state specified by the Bloch vector $\vec{a}$, the probability of obtaining outcome $\ket{0}_D$ is
\begin{equation}
    p = \tr(\rho\ket{0}_D\bra{0}_D)  = \frac12(1+a_3),
\end{equation}
and the probability of obtaining outcome $\ket{1}_D$ is $1-p = \frac12(1-a_3)$. The Fisher information is thus
\begin{equation}\label{eq:fisher}
    \F(\lambda) = \frac{(\partial_\lambda a_3)^2}{1-a_3^2},
\end{equation}
where $a_3$ is the third component of the Bloch vector of the detector's state, given by  \eqref{eq:az}. 

Since temperature is the parameter of interest, we set $\xi$ to be $T$, denoting  the Fisher information for temperature simply by $\F$:
\begin{equation}\label{eq:fisher-t}
    \F = \F(T) = \frac{(\partial_T a_3)^2}{1-a_3^2},
\end{equation}
and shall refer to it simply as the Fisher information. We shall find it convenient to rescale $\F$ by $T^2$ in order to more easily compare for various $T$ . 

\begin{table}[h]
    \centering
    \begin{tabular}{cc}
        Comoving in dS  & Accelerated in AdS  \\
        \hline
        Temperature $T$ & Temperature $T$ \\
        Acceleration $a$ & Curvature $k$ \\
        Energy gap $\Omega$ & Energy gap $\Omega$ \\
        Initial state $\theta$ & Initial state $\theta$ \\
        Interaction time $\tau$ & Interaction time $\tau$ \\
         & Boundary condition $\zeta$
    \end{tabular}
    \caption{List of relevant parameters for temperature estimation in dS$_4$ and AdS$_4$. Acceleration $a$ and curvature $k=1/\ell$ play analogous roles, as explained below. AdS$_4$ has  the chosen boundary condition (see~\eqref{eq:ads-response}) as an extra parameter.}
    \label{tab:params}
\end{table}

In Table \ref{tab:params}, we list the relevant parameters for both dS  and AdS that the Fisher information depends on, writing $k=1/\ell$.
Note that since the temperature in either spacetime is given by $T = \sqrt{a^2 \pm k^2}/2\pi$, we could choose any two of $T,a,$ and $k$ to parametrize the response per unit time $\G$, Eqs.~\eqref{eq:ds-response-a} and \eqref{eq:ads-response}, and the last would be automatically determined. However, since $\F$ is essentially the square of a derivative of $\G$ with respect to $T$, it takes on a slightly different form depending on the choice of parametrization. 
  
In dS, in order to directly compare the Fisher information for trajectories of the comoving observer with those of nonzero acceleration, we find it more useful to choose the acceleration $a$   as one of the parameters. On the other hand, for the AdS  case, if $a$ and $T$ were chosen as the parameters, then $a$ has to be greater than $2\pi T$ in order to ensure $k = \sqrt{a^2 - 4\pi^2T^2}$ is real. While there is nothing wrong with working with $a$ and $T$, we find it more convenient to work with $k$ and $T$, since both can take any positive real value.

\section{Results}\label{ch:results}

We are interested in the dependence of the Fisher information on various parameters so as to determine its efficacy as a diagnostic of the structure of spacetime.  
We will generally illustrate our results in terms of plots of $\F$ as a function of the relevant parameters, particularly the evolution time $\tau$ and the detector gap $\Omega$.

\subsection{de Sitter Spacetime}\label{sec:results-ds}

In this section, we examine the Fisher information of the comoving and accelerated observer in dS. Since we  choose  to work with $a$ and $T$ as parameters, we can use the response per unit time of the accelerated observer, Eq.~\eqref{eq:ds-response-a}, to calculate $\F$, and the comoving case coincides with $a = 0$. 

\begin{figure}[h!]
    \centering
    \includegraphics[width=\textwidth]{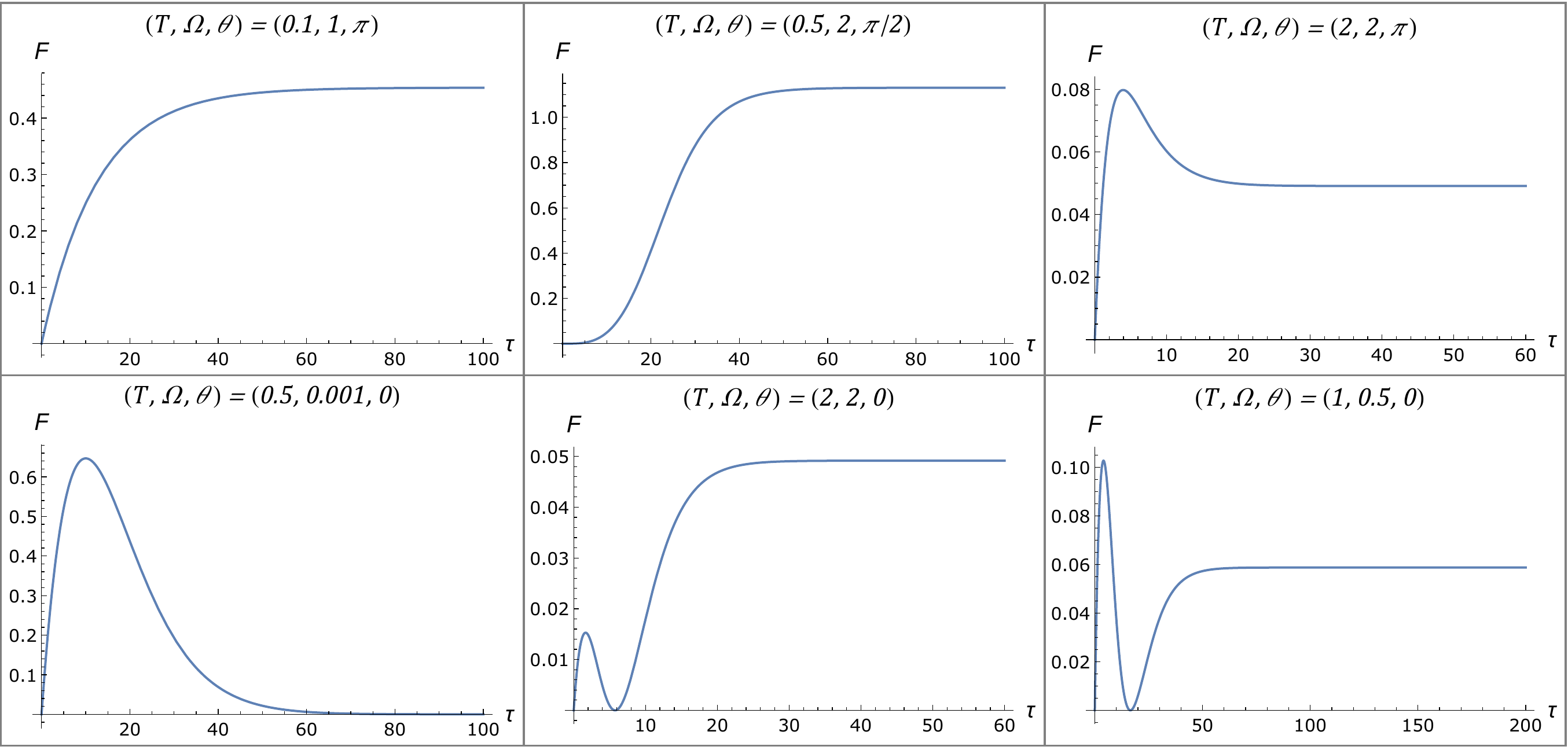}
    \caption{A gallery of different qualitative behaviors of the time evolution of Fisher information in dS$_4$. Only three of them (top row) were previously discovered in Ref. \cite{huang2018quantum}.}
    \label{fig:gallery-ds}
\end{figure}

Let us begin by studying how the Fisher information evolves with time $\tau$ as the detector interacts with the quantum field. Inserting \eqref{eq:az} in  \eqref{eq:fisher-t}, we find several qualitatively distinct behaviors of $\F(\tau)$, representing each type in Fig. \ref{fig:gallery-ds}. Note that all the instances shown in Fig. \ref{fig:gallery-ds} are for the comoving detector, with $a=0$, because we did not find any new  behaviour arising for  nonzero acceleration.

 We see that the Fisher information has at most one local maximum as a function of $\tau$, which may or may not be a global maximum.  If this maximum occurs, it takes place at relatively early times.  
 Of the six types of behaviours for a comoving detector in dS, depicted in Fig. \ref{fig:gallery-ds},
only three   were previously discovered \cite{huang2018quantum}.

\begin{figure}[h!]
    \centering
    \includegraphics[width=\textwidth]{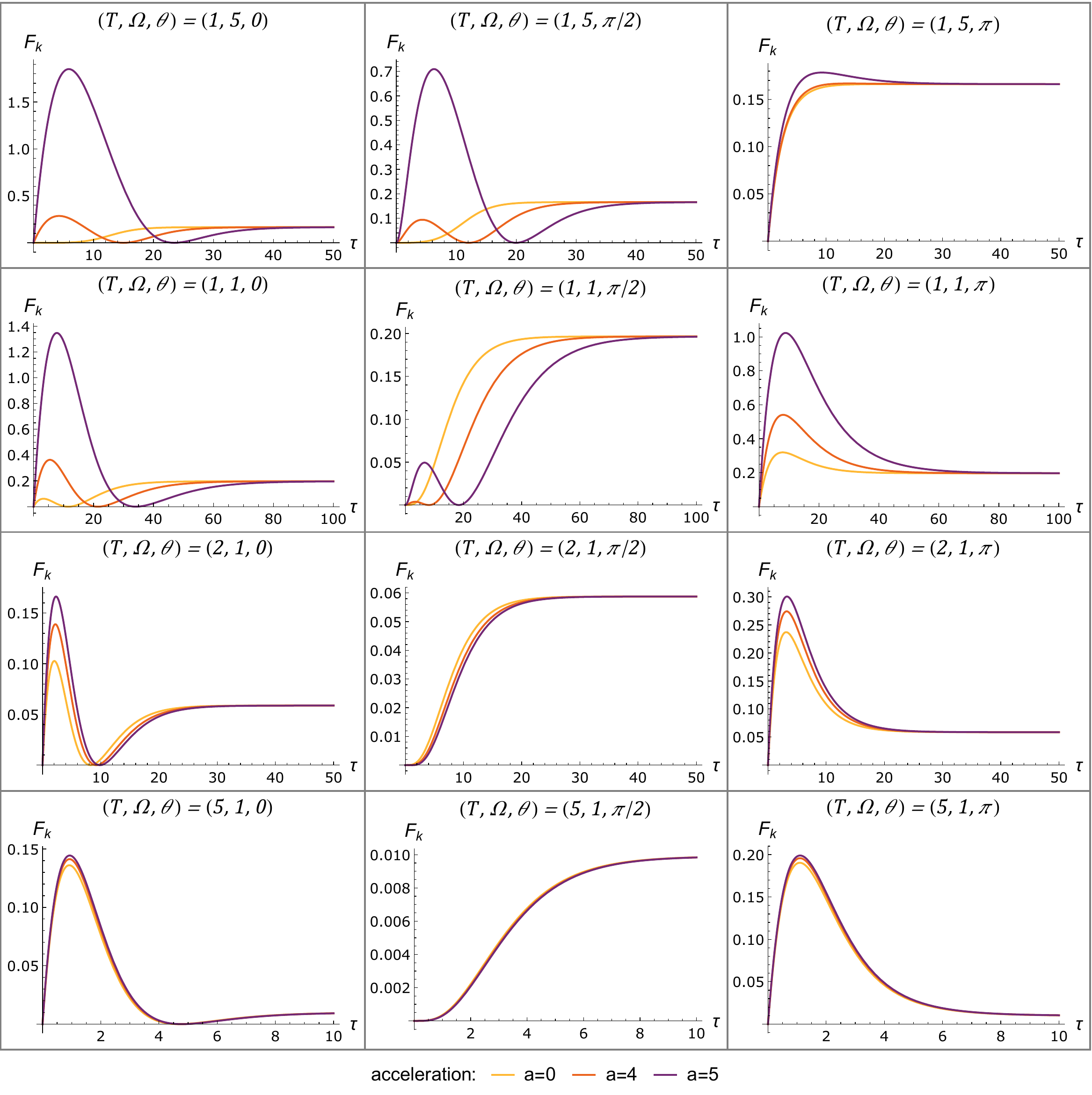}
    \caption{The effect of acceleration on the Fisher information for a detector in dS$_4$.}
    \label{fig:accels}
\end{figure}

If detector acceleration does not give rise to new qualitative behavior, what then is the  effect of acceleration? Fig. \ref{fig:accels} compares the time evolution of $\F$ under different accelerations for a variety of $(T, \Omega, \theta)$ values. Graphs in the same row share the same $(T, \Omega)$ values, starting with $(T, \Omega) = (1,1)$ in the second row, with $T$ increasing below and $\Omega$ increasing above. Each column corresponds to a different initial state: on the left, $\theta = 0$ corresponds to the ground state $\ket{0}$; on the right, $\theta = \pi$ corresponds to the excited state $\ket{1}$; in the middle, $\theta = \pi/2$ corresponds to an equal superposition of ground and excited states $(\ket{0} + \ket{1})/\sqrt{2}$. 

In general, acceleration can change the type of behavior of the Fisher information. For example, a local maximum can become a global maximum, as acceleration increases. For larger $T$, the effects of different accelerations become less distinguishable.

One clear feature  in Figure~\ref{fig:accels} is that for all choices of $(T, \Omega, \theta)$, the Fisher information asymptotes to a fixed value at large $\tau$, which  is the proper time parametrizing the interaction of the detector with the environment. Intuitively, the Fisher information tends to an equilibrium value as the detector thermalizes with the environment, at a time $\tau$ much larger than the timescale of the UDW transition.

Analytically, it is easy to take the $\tau \to \infty$ limit of Eq. \eqref{eq:az}:
\begin{equation}
    \lim_{\tau \to \infty} a_3 = - R,
\end{equation}
and therefore
\begin{equation}
    \lim_{\tau \to \infty} \F = \lim_{\tau \to \infty} \frac{(\partial_T a_3)^2}{1-a_3^2} = \frac{(\partial_T R)^2}{1-R^2}.
\end{equation}
Using \eqref{eq:A} and \eqref{eq:B}, it is straightforward to show that
\begin{equation}
    R^{\text{dS}} = -\tanh\left( \frac{\Omega}{2T} \right)
\end{equation}
which is a consequence of the KMS relation. We see that $R$ does not depend on the acceleration $a$, and so
 the asymptotic Fisher information $\F_{\text{asym}}$ 
\begin{equation}\label{eq:asym}
    \F_{\text{asym}}^{\text{dS}} = \frac{\Omega^2}{4T^2} \,\sech^2\left(\frac{\Omega}{2T}\right)
\end{equation}
is also identical \emph{for all accelerations and initial states}.
Since $\F_{\text{asym}}$ is a function of $\Omega$ and $T$, we conclude that the asymptotic Fisher information is a thermal property in dS. In the next section, we will see that this remains true for accelerated observers in AdS$_4$ with any boundary condition. In fact, the asymptotic Fisher information for \emph{any} observer in either dS and AdS is identical, given by Eq.~\ref{eq:asym}. 

\begin{figure}[h!]
    \centering
    \includegraphics[width=\textwidth]{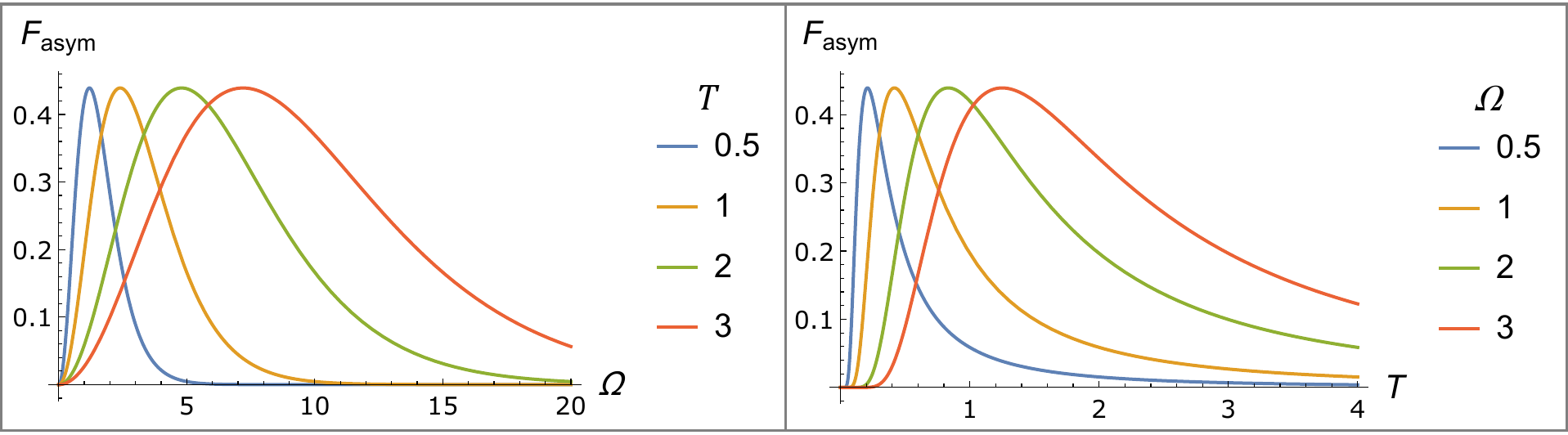}
    \caption{The asymptotic Fisher information as a function of energy gap $\Omega$ (left) and temperature $T$ (right). Note that there is always one $\Omega$ ($T$) that maximizes the asymptotic Fisher information for fixed $T$ (fixed $\Omega$).}
    \label{fig:asym}
\end{figure}

Fig. \ref{fig:asym} shows how $\F_\text{asym}$ qualitatively depends on temperature $T$ and energy gap $\Omega$. We see that for each value of $T$ ($\Omega$), $\F_\text{asym}$ achieves maximum for some value of $\Omega_\text{max}$ ($T_\text{max}$), and the value of $\Omega_\text{max}$ ($T_\text{max}$) is greater for higher $T$ ($\Omega$).

Thanks to the simple form of $\F_\text{asym}$, we can in fact analyze its quantitative properties. For fixed $T$, we can easily find the $\Omega$ that maximizes $\F_{\text{asym}}$:
\begin{equation}
    \D{\F_\text{asym}}{\Omega} = \frac{\Omega}{4T^3} \sech^2\left( \frac{\Omega}{2T} \right) \left[ 2T - \Omega \tanh\left( \frac{\Omega}{2T} \right) \right] = 0.
\end{equation}
Similarly, we can find the $T$ that maximizes the asymptotic Fisher information for fixed $\Omega$:
\begin{equation}
    \D{\F_\text{asym}}{T} = \frac{\Omega^2}{4T^4} \sech^2\left( \frac{\Omega}{2T} \right) \left[ -2T + \Omega \tanh\left( \frac{\Omega}{2T} \right) \right] = 0.
\end{equation}
It is easy to see they both have the same solution  
\begin{equation}
    \tanh\left( \frac{\Omega}{2T} \right) = \frac{2T}{\Omega}
\end{equation}
implying 
\begin{equation}\label{eq:omega-max}
    \Omega_\text{max} = 2x_0T,
\end{equation}
where   $x_0\tanh x_0 = 1$ (or  $x_0 = 1.1996786 \approx 1.200$). The temperature $T_\text{max}$ that maximizes $\F_\text{asym}$ for fixed $\Omega$ is
\begin{equation}
    T_\text{max} = \frac{\Omega}{2x_0}.
\end{equation}


We also see from  Figure~\ref{fig:accels}, that for some values of $(T, \Omega, \theta, a)$, the asymptotic Fisher information is the maximal $\F$ achieved in the course of the time evolution (e.g. all three accelerations with $(T, \Omega, \theta) = (1,1,\pi/2)$). However, for other values of $(T, \Omega, \theta, a)$ (e.g. all three accelerations with $(T, \Omega, \theta) = (1,1,\pi)$), $\F$ reaches its global maximum at some intermediate $\tau$, and then asymptotes to a smaller value. In order to achieve maximal precision in the estimation, it is therefore valuable to consider strategies that involve letting the detector interact with the field for some time $\tau$ at which the \emph{temporal maximum} of Fisher information is achieved, and then measuring the detector immediately. 

\begin{figure}[h!]
    \centering
    \includegraphics[width=\textwidth]{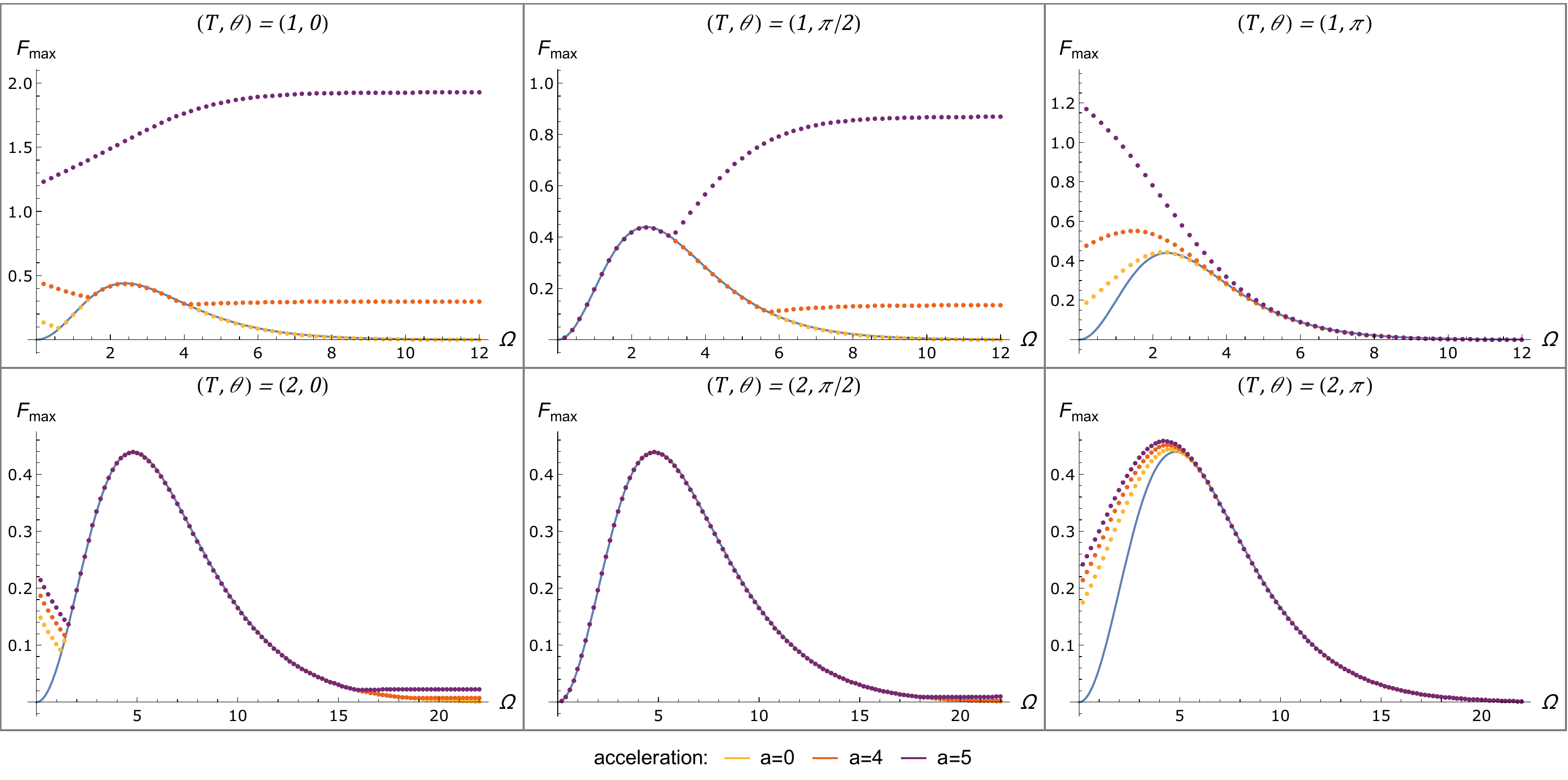}
    \caption{(Approximate) temporal maximum of the Fisher information in dS. Every panel corresponds to a different choice of $(T,\theta)$. Again, the three colors correspond to different accelerations. The asymptotic Fisher information computed in the previous section is also shown in a solid curve.}
    \label{fig:temporalmax-ds}
\end{figure}

How much enhancement in the Fisher information do we see if we pursue this estimation strategy? To find out, 
we approximate the temporal maximum with the maximum of a list of values of $\F$ evaluated at discrete times $\tau$ for various values of $\Omega$.   We refer to the approximate temporal maximum of $\F$ as $\F_\text{max}$, and plot in Fig. \ref{fig:temporalmax-ds}  $\F_\text{max}$ for fixed $(T,\theta)$ and different accelerations as a function of detector energy gap $\Omega$.

There are a few things we can observe from Figure \ref{fig:temporalmax-ds}. First, when the temporal maximum is greater than  the asymptotic value,
higher acceleration exaggerates this difference.
Second, there appear to be two different regimes: one where the temporal maximum is greater than the asymptote for small $\Omega$ (e.g. all of the $T=2$ examples and $(T,\theta) = (1,\pi)$), and the other where the temporal maximum is greater at large $\Omega$ (e.g. $(T,\theta) = (1,0)$ and $(T,\theta) = (1,\pi/2)$.) While the mechanism of this enhancement is not yet understood, we see that considering the estimation strategy of only allowing the detector to interact with the field for a finite amount of time gives significant enhancement in $\F$ for a wide range of parameters in dS. This result demonstrates the power of the open quantum systems method, since this enhancement is only visible when the full dynamics of the detector is taken into account.

\subsection{Anti-de Sitter Spacetime}

\begin{figure}[h!]
    \centering
    \includegraphics[width=\textwidth]{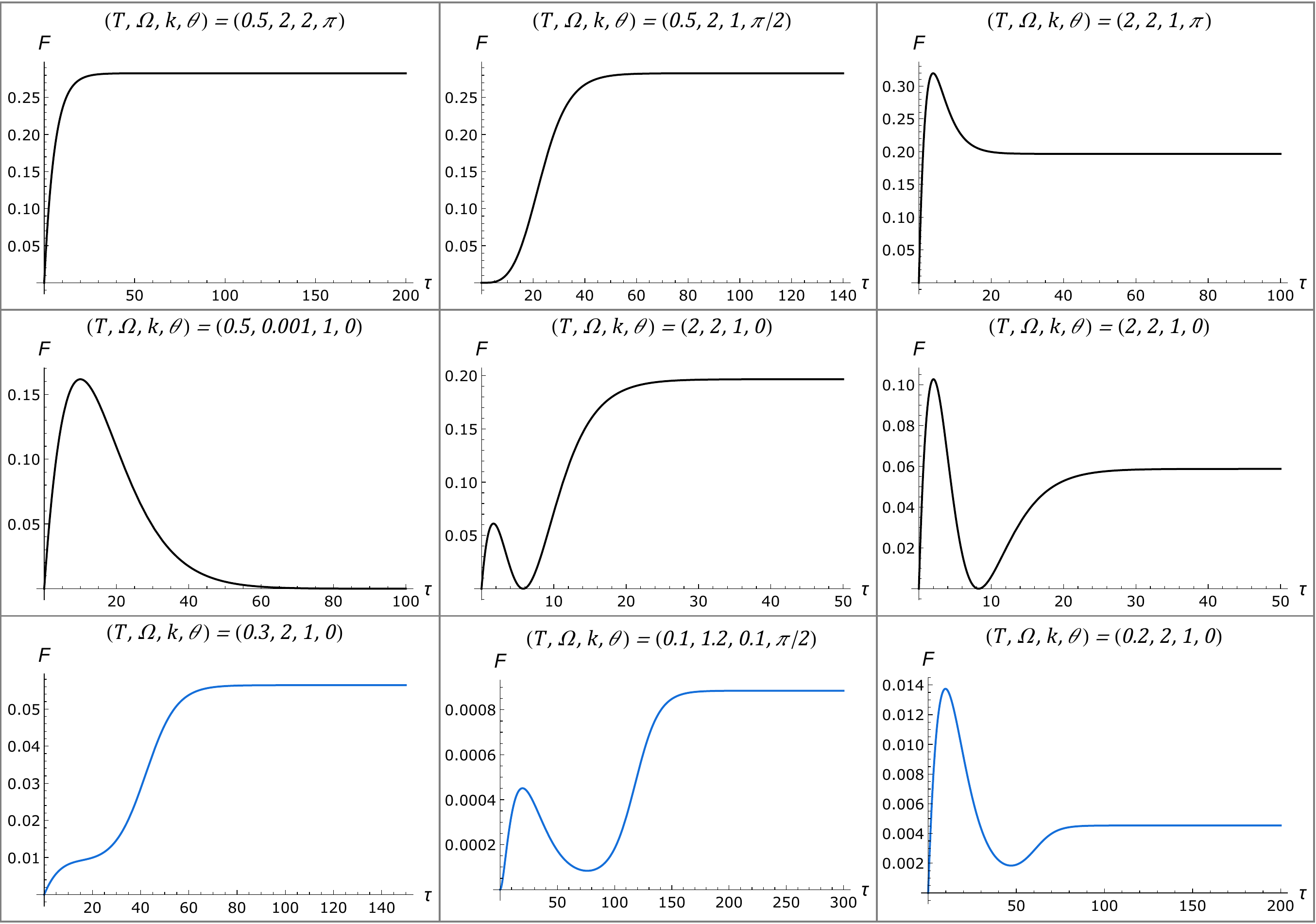}
    \caption{A gallery of different qualitative behaviors of the time evolution of Fisher information in AdS$_4$. The curves in black are under transparent boundary condition ($\zeta = 0$), and the curves in blue are under Dirichlet boundary condition ($\zeta = 1$). The first six panels (aka the ones with transparent boundary condition) are qualitatively identical to those in Fig. \ref{fig:gallery-ds}. We see that AdS$_4$ encompasses all the qualitative behaviors present in dS$_4$, as well as three (bottom row) that are not present in dS$_4$, under nontrivial boundary conditions.}
    \label{fig:gallery-ads}

\end{figure}

As before, we begin by examining the different qualitative behaviors of $\F(\tau)$ in AdS, shown in Figure \ref{fig:gallery-ads}. We identiy nine qualitatively distinct behaviours across all three boundary conditions in AdS, among which the first six are identical to the dS case.  This is unsurprising because the response in AdS  with transparent boundary condition ($\zeta = 0$) is \emph{identical} to that in dS  (for the same temperature and acceleration/curvature). Hence we should expect the transparent AdS case to exactly reproduce the dS behaviors.

However, we find that the extra degrees of freedom added by the two nontrivial boundary conditions ($\zeta = \pm 1$) enable new types of behavior, shown in the last row of Figure \ref{fig:gallery-ads}. The three cases shown are all for Dirichlet boundary condition ($\zeta = 1$), but these behaviors are present for Neumann boundary conditions as well.

\begin{figure}[h!]
    \centering
    \includegraphics[width=\textwidth]{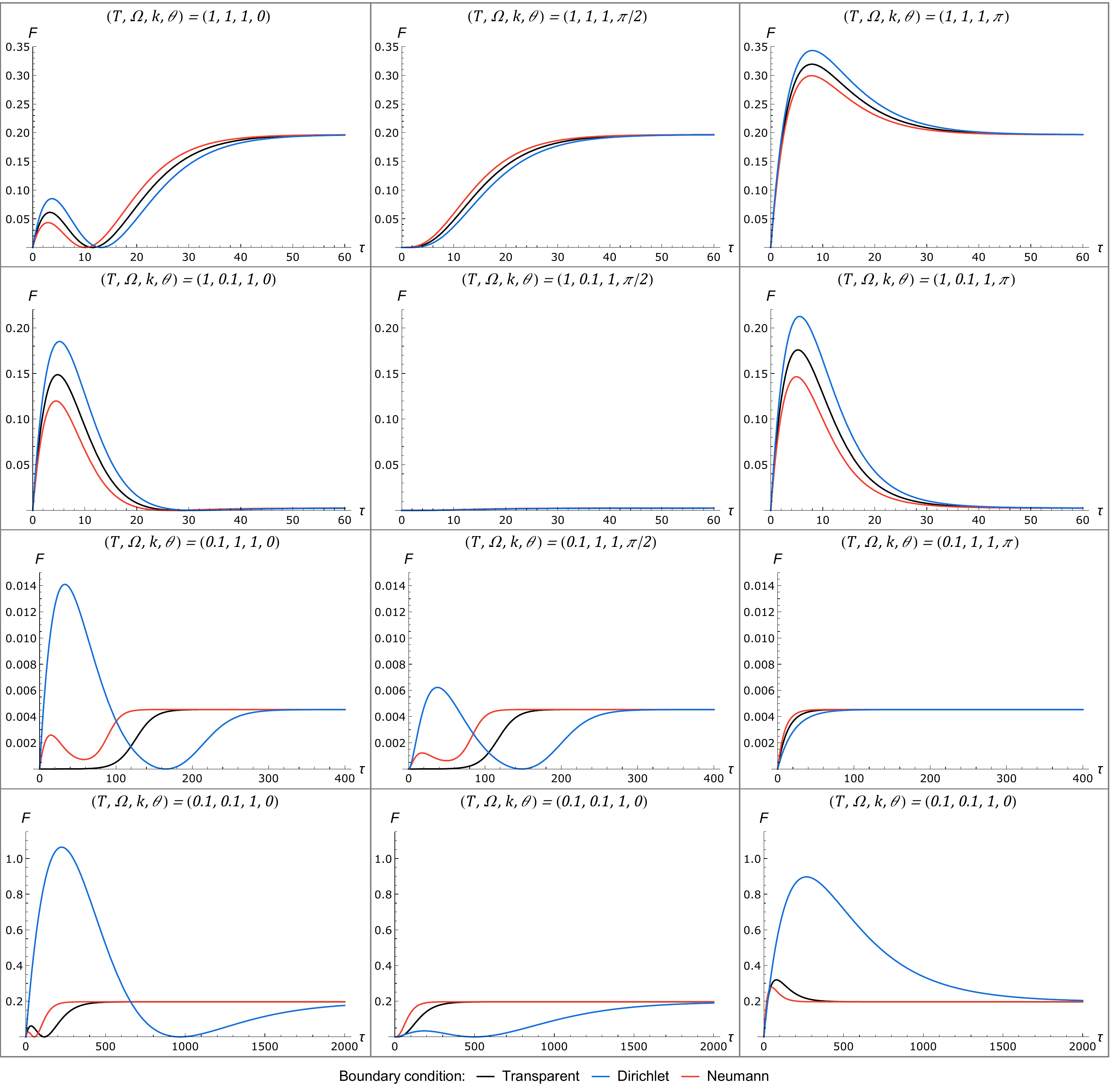}
    \caption{The effect of different boundary conditions on the Fisher information in AdS$_4$.}
    \label{fig:bcs}

\end{figure}

It is  natural to directly compare the behavior of the Fisher information under different boundary conditions. Figure \ref{fig:bcs} shows the time evolution of $\F$ under the three boundary conditions for a variety of $(T, \Omega, \theta)$ values. Recall that $\zeta = 0$ corresponds to transparent boundary conditions, and $\zeta = \pm 1$ correspond to Dirichlet and Neumann boundary conditions respectively. We see that while the transparent case is always ``in between'' the other two cases in some sense, and the change in boundary condition gives rise to drastic changes in behavior for some $(T, \Omega, k, \theta)$ values (e.g. all the $(T,\Omega) = (0.1,0.1)$ examples.)

One might suspect similar asymptotic behavior for AdS  as for dS.  Indeed, the asymptotic Fisher information in AdS  for all three boundary conditions is \emph{exactly the same} as that in dS. This is because the asymptotic Fisher information only depends on $R$, and
\begin{equation}
    R^\text{AdS} = R^\text{dS} = - \tanh \left( \frac{\Omega}{2T} \right),
\end{equation}
so the exact analysis from the previous section follows. Importantly, this tells us that the asymptotic Fisher information only depends on temperature $T$ and energy gap $\Omega$, for all observers in dS and AdS, and is unaffected by the details of their trajectories.

\begin{figure}[h!]
    \centering
    \includegraphics[width=\textwidth]{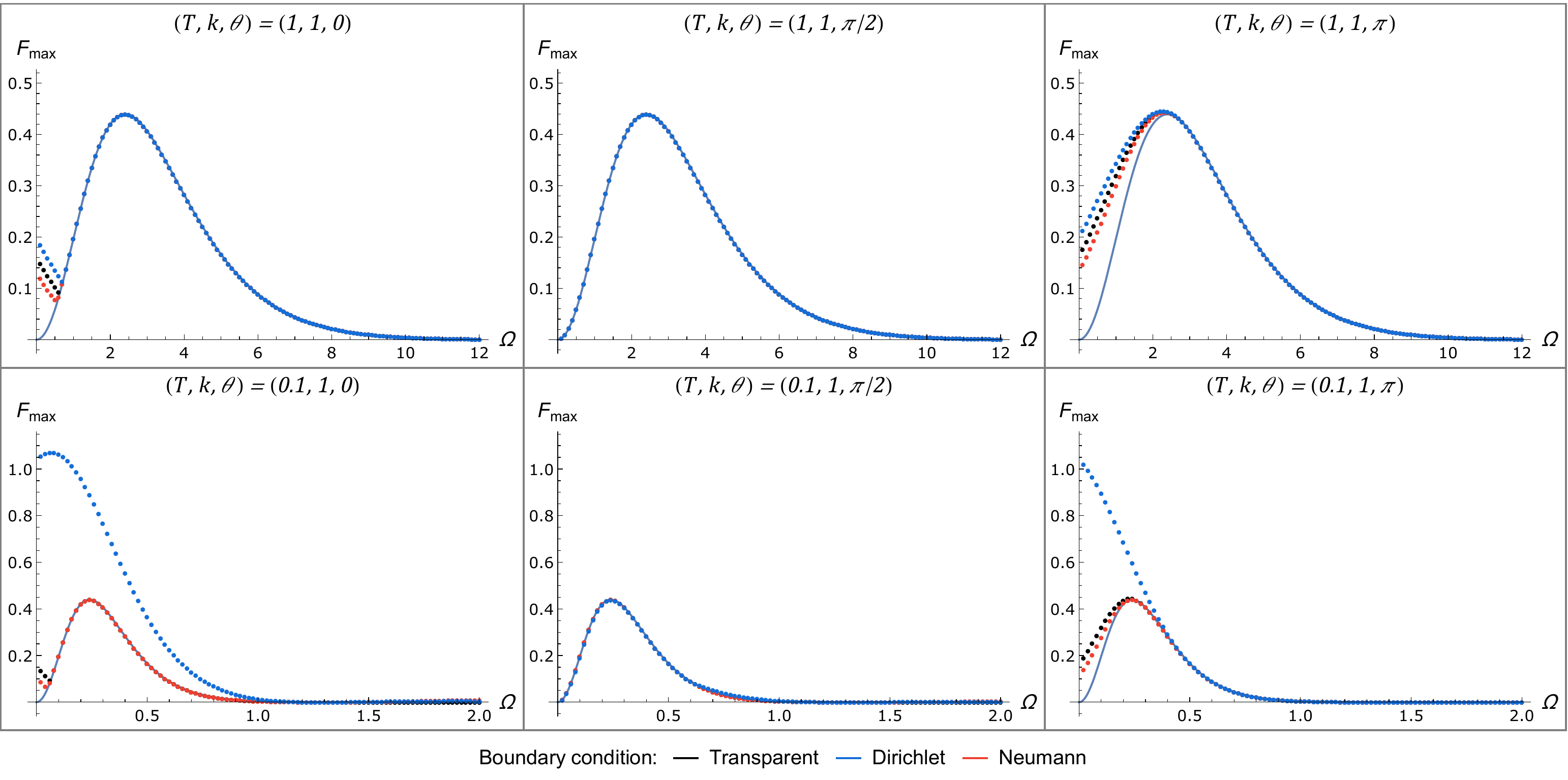}
    \caption{(Approximate) temporal maximum of the Fisher information in AdS$_4$. Every panel corresponds to a different choice of $(T, k, \theta)$. Again, the three colors correspond to different boundary conditions. The asymptotic Fisher information is shown as a solid curve.}
    \label{fig:temporalmax-ads}
\end{figure}

Finally, let us again consider the estimation strategy of only allowing the detector to interact with the quantum field for a finite amount of time, and try to capture the temporal maximum achieved in the course of the time evolution. The approximate temporal maxima are computed via the same method as described in Section \ref{sec:results-ds}, and are shown in Figure \ref{fig:temporalmax-ads}.

As before, we observe significant enhancement for a range of parameters, while our theoretical understanding of the mechanism of such enhancement remains limited. In contrast with the dS  case, at least for the range of parameter values explored, we see enhancement in AdS  only for small $\Omega$ values. We note that there appears to be little to no enhancement if the initial state is the superposition state ($\theta = \pi/2$).

\section{Conclusion}\label{ch:conclusion}



We have studied the Fisher information for estimating the temperature experienced by detectors of various trajectories in dS  and AdS spacetimes. The following general features emerged.
 
Our results in part extend an earlier study of Fisher information for comoving detectors in dS
\cite{huang2018quantum}. We find that these previous results  are not sufficiently representative of the behavior of the Fisher information in dS: we identify three additional types of behavior for the Fisher information in dS. 
 
 If a comoving (zero-acceleration) detector in dS  and an accelerated detector in AdS  with \emph{transparent} boundary condition experience the same temperature, then they have the exact same response function, and therefore the same Fisher information.
    
 For long interaction times that allow the detector to thermalize with the environment,   a detector on \emph{any} trajectory in either spacetime gives the same Fisher information. Furthermore, the energy gap $\Omega$ that maximizes the Fisher information given temperature $T$ can be analytically calculated, and is given by Eq. \eqref{eq:omega-max}.

 For a wide range of parameter settings, the Fisher information can be enhanced by restricting the interaction time $\tau$ to some intermediate value, and catching the temporal maximum before the detector thermalizes with the environment. We see two distinct regimes for this enhancement in dS, whereas in AdS  this enhancement only appears possible for small $\Omega$.

Many directions are open for future work.

While our detailed study of the temporal maximum of the Fisher information, has allowed us to identify strategies that maximize the Fisher information, 
the parameter space is large, and so it  is by no means clear as to how to determine the optimal estimation strategy.  
 An immediate next task would be to find the \emph{global} maximum of the Fisher information in the entire parameter space, and therefore identifying \emph{the} optimal estimation strategy.

 The framework we have employed  can be easily generalized to many related problems. For example, one could  extend our calculations to the multivariate case using the \emph{Fisher matrix} to  capture the effects of correlations between the parameters. It is also possible to   extend our estimation strategy to include initializing the detector in a mixed initial state. Since a mixed state can be regarded as part of a larger, entangled system, considering mixed states would give us insight as to whether entanglement is a resource for this metrological task. 

Furthermore, there are many other choices of projective measurements as well as POVM measurements that can be made on the detector's quantum state, and the Fisher information clearly depends on the choice of measurement.  For the cases examined in this work, projective measurement in the computational basis \cite{huang2018quantum} in fact yields the maximal Fisher information across all choices of measurement. This in fact is the \emph{quantum Fisher information}  \cite{petz2011introduction, giovannetti2006quantum, paris2009quantum}. An obvious next step is to investigate the quantum Fisher information in the context of our setting to verify (or refute) the claim of Ref. \cite{huang2018quantum}.

Besides considering other estimation strategies, our framework can also be easily generalized to other metrological tasks.   For example, we could take $k=1/\ell$ in AdS to be the parameter of interest $\xi$ in Eq. \eqref{eq:fisher} and compute the Fisher information to estimate $k$. 
The same could be done for various properties of the detector such as the energy gap $\Omega$. 

Finally, a study relativistic quantum metrology in other spacetimes could prove to be fruitful. For example, the BTZ black hole \cite{banados1992black} in $(2+1)$-dimensional AdS admits   a Wightman function of similar form to the vacuum Wightman function studied here \cite{lifschytz1994scalar}, and is known to have interesting vacuum entanglement properties \cite{Henderson:2017yuv,Robbins:2020jca}. It would be interesting to understand how the Fisher information for estimating the black hole mass behaves, and how  the presence of an event horizon affects it.  Likewise, given previous results indicating  that topology plays a role in the entanglement structure of spacetimes \cite{martin2016spacetime},  it would be interesting to investigate its effect on Fisher information and metrology.

\section*{Acknowledgements}

This work is based on an essay completed as part of the Perimeter Scholars International program.
This work was supported in part by the Natural Sciences and Engineering Research Council of Canada.




\bibliographystyle{./style_files/JHEP}
\bibliography{references}






\end{document}